\documentclass[%
reprint,%
aps,%
pra,%
groupedaddress,%
superscriptaddress,%
assume,%
amsmath,%
floatfix
]{revtex4-2}

\usepackage{graphicx}
\usepackage[
  colorlinks=true,
  urlcolor=blue,
  linkcolor=blue,
  citecolor=blue
]{hyperref}
\usepackage{siunitx}
\usepackage{braket}
\usepackage{gensymb}
\usepackage{amssymb}

\usepackage{layouts}
\usepackage{ulem}

\begin{document}
\title{On-Chip Optical Skyrmionic Beam Generators}

\author{Wenbo Lin}
\email{lin.w.ab@m.titech.ac.jp}
\email{lin.w.ebb4@m.isct.ac.jp}
\affiliation{
Institute of Innovative Research, Tokyo Institute of Technology, 2-12-1 Ookayama, Merugo, Tokyo 152-8550, Japan
}
\author{Yasutomo Ota}
\affiliation{
Department of Applied Physics and Physico-Informatics, Faculty of Science and Technology, Keio University, 3-14-1 Hiyoshi, Kohoku-ku, Yokohama, Kanagawa 223-8522, Japan
}
\author{Yasuhiko Arakawa}
\affiliation{
Institute for Nano Quantum Information Electronics, The University of Tokyo, 4-6-1 Komaba, Meguro, Tokyo 153-8505, Japan
}
\author{Satoshi Iwamoto}
\affiliation{
Research Center for Advanced Science and Technology, The University of Tokyo, 4-6-1 Komaba, Meguro, Tokyo 153-8904, Japan
}
\affiliation{
Institute of Industrial Science, The University of Tokyo, 4-6-1 Komaba, Meguro, Tokyo 153-8505, Japan
}

\date{\today}

\begin{abstract}
Optical skyrmion beams, which encompass two-dimensional topology in their spatial structures, are promising for ultra-dense optical communications and advanced matter manipulation. Generating such light beams via a chip-based approach will vastly broaden their applications and promote the advancement of untapped fundamental science. Here, we present a breakthrough in chip-based technology by experimentally demonstrating on-chip devices capable of generating optical skyrmions with tailored topological invariants. These devices, fabricated with high precision, exhibit behavior that closely aligns with theoretical predictions and numerical simulations. The realization of on-chip optical skyrmion beam generators ushers a new dawn in optical and material science.
\end{abstract}

\maketitle

\section{Introduction}
Topology plays a crucial role in revealing the intricate physics underlying optical phenomena. This powerful tool has been instrumental in classifying a diverse array of unique optical structures that manifest in real~\cite{nye1983polarization,allen1992orbital,soskin1997topological,shen2019optical} and momentum space~\cite{haldane2008possible,wang2009observation,lu2014topological}, space-time~\cite{chong2020generation,alonso2020complete,shen2023roadmap}, and even synthetic spaces~\cite{lustig2019photonic,lustig2021topological}, thereby broadening their applications. For instance, vortices manifested in the spatial structure of phase and polarization states of a light beam are characterized by topological invariants and thus are well-conserved physical entities, promising high-capacity and robust telecommunications~\cite{yan2014high,zhao15high}.

As our understanding of the topological aspect of light deepens, higher-dimensional and more sophisticated topologies beyond these simple one-dimensional vortices are gradually being elucidated. A prominent example is the two-dimensional (or baby-) skyrmion~\cite{shen2024optical}. A (2-D) skyrmion is a non-trivial structure of a field of normalized three-vectors (pseudospins) in two dimensions. The vector field on a sphere (or on a plane $\mathbb{R}^2$, which can be mapped to a sphere $S^2$ via stereographic projection) must wrap around the unit sphere an integer number of times. This configuration is classified by the second homotopy group of spheres $\pi_2\left(S^2\right) = \mathbb{Z}$. The wrapping number, known as the skyrmion number $N_\mathrm{sk}$, serves as the topological invariant characterizing a skyrmion. In optical fields, such topological structures are recently discovered in the spatial distributions of electric/magnetic field vectors~\cite{tsesses2018optical,davis2020ultrafast,shen2021supertoroidal}, optical spin vectors~\cite{du2019deep,dai2020plasmonic,lei2021photonic,lin2021photonic,zhang2021bloch}, and Stokes vectors~\cite{donati2016twist,guo2020meron,gao2020paraxial,lin2021microcavity,shen2022generation,luo2023non,marco2024propagation,li2024realization}. The unique topology of optical skyrmions imparts extraordinary spatial properties and rich dynamics, leading to potential applications in single-shot Mueller matrix polarimetry~\cite{suarez2019mueller}, pico-metric sensing~\cite{yang2023spin}, novel optical tweezers~\cite{wang12optical}, exotic laser processing~\cite{zhu15transverse,tamura2024direct}, and solid-state topological excitation manipulation~\cite{donati2016twist}. Similar to their one-dimensional counterparts, optical skyrmions are anticipated to be robust carriers of information~\cite{liu2022disorder} and hence may be useful in optical communications and quantum information processing.

Despite recent advancements in optical skyrmions, most current methods for generating light beams with skyrmionic topology rely on bulk optics~\cite{beckley2010full,galvez12poincare,shen2021topological,shen2021supertoroidal,shen2022generation,sugic2021particle,zhang2022optical,ehrmanntraut2023optical,shen2023topological,kerridge2024optical,marco2024propagation,srinivasa2024optical}. Exploration within integrated photonics remains limited~\cite{lin2021microcavity}, with no experimental demonstrations to date. With its compactness, on-chip generation technology can mitigate the influence of external disturbances such as mechanical vibrations, which may compromise beam stability or cause fluctuations in the beam’s position relative to manipulation targets, thus revolutionizing many of the aforementioned applications and potentially spurring unseen ones. For example, highly integrated skyrmion light sources may result in ultra-dense and robust optical communications. Moreover, the synergy between on-chip generation and nanofabrication technology, which is suitable for mass production, makes chip-based technology highly demanded for the widespread adoption of optical skyrmion beams.

In this study, we demonstrate on-chip optical skyrmion beam generators based on the silicon photonics platform. Silicon microrings with double diffraction gratings are employed to tailor the spin and orbital angular momentum (OAM) of light, synthesizing all possible polarization states on the unit Poincar\'{e} sphere, thereby fulfilling the skyrmion topology. Precisely fabricated devices on the nanometer order closely follow the predictions of numerical simulations, and these results are well-represented by expanding the theoretical model previously proposed~\cite{lin2021microcavity}. We also show that the skyrmion number can be controlled by a simple structural modification of the device, enabling high-dense integration of skyrmion beam generators with various topological charges. Our innovative skyrmionic beam generator could pave the way for novel laser devices, sensor technologies, optical telecommunications, and applications in atomic or solid-state physics within a chip-based framework.

\section{Device structure and operating mechanism}
\begin{figure*}[htbp]
    \centering
    \includegraphics{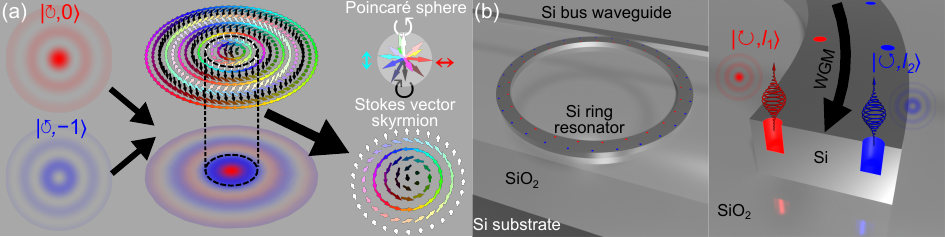}
    \caption{
        (a)~A sketch depicting the mechanism of optical skyrmion beam generation. The superposition of two optical modes with orthogonal polarization states and different OAMs ($\ket{\circlearrowleft,0}$ and $\ket{\circlearrowright,-1}$ are exemplified here) yields a spatially-varying polarization field (visualized by Stokes vectors in the central panel). Near the center of the mode, an iso-polarization circle emerges ($\ket{\circlearrowright}$ in this example, indicated by the black dashed line), and all polarization states represented on the Poincar\'{e} sphere appear inside the circle, sculpting a Stokes skyrmion.
        (b)~Schematics depict an optical skyrmion beam generator. The base structure is a silicon ring resonator fabricated on a silicon-on-insulator substrate (left panel). A double angular diffraction grating consisting of half-etched holes is arranged on the ring (right panel). Two angular gratings are designed to diffract distinct polarization states and OAM orders.
    }
    \label{fig:Concept}
\end{figure*}

Figure~\ref{fig:Concept}(a) schematically illustrates the generation mechanism of an optical skyrmionic beam. The skyrmion pseudospin texture manifests in the spatial mode cross-section of a superposition beam composed of two light beams with orthogonal polarization states and different OAM orders~\cite{lin2021microcavity}. Two light beams with $\ket{\circlearrowright,0}$ and $\ket{\circlearrowleft,-1}$ are exemplified in the left panel of Fig.~\ref{fig:Concept}(a). Here, $\ket{\circlearrowright,l}$ denotes $\ket{\circlearrowright}$-polarized state with an OAM order $l$, where $\ket{\circlearrowright}$ and $\ket{\circlearrowleft}$ represent right- and left-handed circular polarization. By superimposing these two light beams, the spatial polarization state varies across the superposition beam cross-section due to differences in the spatial phase distribution originating from the OAM difference, as presented in the central panel of the same figure. Throughout this paper, the out-of-plane component ($S_3/S_0$) of the pseudospin vector $\boldsymbol{s} = \left(S_1, S_2, S_3 \right)/S_0$, which is the normalized Stokes vector, is indicated by saturation/brightness, whereas in-plane azimuth $\mathrm{arctan}{\left(S_1/S_2\right)}$ is mapped to hue, as depicted with the unit Poincar\'{e} sphere at the top-right corner of Fig.~\ref{fig:Concept}(a). Typically, OAM order affects the spatial profile of the intensity, with the component having a lower absolute value of OAM order being predominant at the center of the beam. Thus, in this example, the $\ket{\circlearrowright}$-polarized component with a zero-order OAM (spin-down, indicated by black arrows) predominates at the beam center. If the $\ket{\circlearrowright}$ component has an intensity-null closed loop around the beam center (or $\ket{\circlearrowleft}$ becomes dominant at infinity), a closed loop of $\ket{\circlearrowleft}$ (spin-up, shown by white arrows) forms around the center. Here, spatial profiles of beams are assumed to be defined by the $\left|l\right|$-th Bessel function of the first kind $J_{\left|l\right|}$, for consistency with the discussion afterward. Consequently, a circle of spin-up emerges at the radial position corresponding to the first zero of $J_0$ that characterizes the spatial profile of the $\ket{\circlearrowright,0}$ component. Within the domain enclosed by this circle $\Omega$, the pseudospin flips upward while swirling, and all polarization states represented by the unit Poincar\'{e} sphere emerge, thus fulfilling the skyrmion topology. The topological invariant can be computed by evaluating the skyrmion number density
\begin{equation}
    dN_\mathrm{sk} = \frac{1}{4\pi} \boldsymbol{s}\cdot\left(\partial{_X}\boldsymbol{s}\times\partial{_Y}\boldsymbol{s}\right)
    \label{eq:1}
\end{equation}
and then integrate it over the domain $N_\mathrm{sk} = \int_{\Omega}{dN_\mathrm{sk}\mathrm{d}X\mathrm{d}Y}$. Here, the skyrmion number density corresponds to the real-space Berry curvature
\begin{equation}
    B = -i\nabla \times \left( \frac{\boldsymbol{E_\parallel}^\ast}{\left| E_\parallel \right|} \nabla \frac{\boldsymbol{E_\parallel}}{\left| E_\parallel \right|} \right)
\end{equation}
divided by $2 \pi$, i.e., $dN_\mathrm{sk} = B/2\pi$, where $\boldsymbol{E}_\parallel$ is the transverse electric field vector on the surface defines the two-dimensional skyrmion, i.e., the electric field in the XY plane. The in-plane component of the pseudospin swirls $\Delta{l} = l_2 - l_1$ times around the beam center, the difference in the OAM orders, hence the Poincar\'{e} sphere is covered  $\left|\Delta{l}\right|$ times, resulting in $N_\mathrm{sk} = \Delta{l}$.

To superimpose two light beams with orthogonal polarization states and distinct OAMs on-chip, a ring resonator device based on the discussions in \cite{lin2021microcavity} is employed, as depicted in Fig.~\ref{fig:Concept}(b). The ring is made of silicon for future integration into silicon photonic systems and is placed on an SiO$_2$ substrate, considering the use of silicon-on-insulator wafers. Two arrays of partially etched holes are arranged on the silicon ring to form a double angular grating. One grating is offset toward the inner sidewall of the ring waveguide, while the other is offset toward the outer sidewall. Due to the spin-orbit interaction of light~\cite{coles2016chirality}, these positional offsets cause the former grating to diffract $\ket{\circlearrowright}$-polarized light and the latter to diffract $\ket{\circlearrowleft}$-polarized light when a clockwise fundamental transverse electric (TE)-like whispering-gallery mode (WGM) is driven, as illustrated in the right panel of Fig.~\ref{fig:Concept}(b). For counterclockwise modes, the handedness of the circular polarizations is reversed. The light diffracted by each grating also carries an OAM, denoted by orders $l_1$ and $l_2$ respectively for the inner and outer gratings, satisfying the conservation of angular momentum equation $l+s=\gamma\left(m-ng\right)$, where $s$ represents the spin angular momentum ($+1$ for $\ket{\circlearrowleft}$ and $-1$ for $\ket{\circlearrowright}$), $\gamma$ is the handedness of the WGM ($-1$ for clockwise and $+1$ for counterclockwise mode), $m$ is the azimuthal order of the WGM, $g$ is the number of grating periods, and $n$ is the diffraction order. When $g$ is close to $m$, $n=1$ can be assumed because higher-order diffractions are not emitted from the ring resonator due to being primarily below the light line. The far-field diffraction profile is then characterized by the $\left| l \right|$-th Bessel function of the first kind (see supplemental materials of \cite{lin2021microcavity} for details). The sign of the OAM order $l = \gamma\left( m - g \right) - s$ is flipped when the base WGM changes from clockwise to counterclockwise.

A silicon-on-insulator wafer with a silicon layer thickness of \SI{220}{\nano\meter} and a buried oxide layer of \SI{3}{\micro\meter} is used for device fabrication. The ring resonator features a radius of \SI{3}{\micro\meter} and a waveguide width of \SI{450}{\nano\meter}. Finite-difference time-domain (FDTD) simulations indicate that a TE-like WGM with an azimuthal order of $m=28$ is present within the telecommunication C-band. For unperturbed ring resonators, numerical simulations reveal the formation of two chiral lines, where local polarization of the cavity mode is circularly polarized due to optical spin-orbit coupling, located approximately \SI{150}{\nano\meter} from the center of the ring waveguide to the inner and outer sidewalls. Angular gratings are aligned along these chiral lines to diffract circularly-polarized light. The diameter of the nanoholes, which are the grating elements, is set to \SI{60}{\nano\meter} to ensure a satisfactory yield. Additional details on the device fabrication process are provided in the Supplemental Material (SM).

\section{Optical characterization}
\begin{figure*}[htbp]
    \centering
    \includegraphics{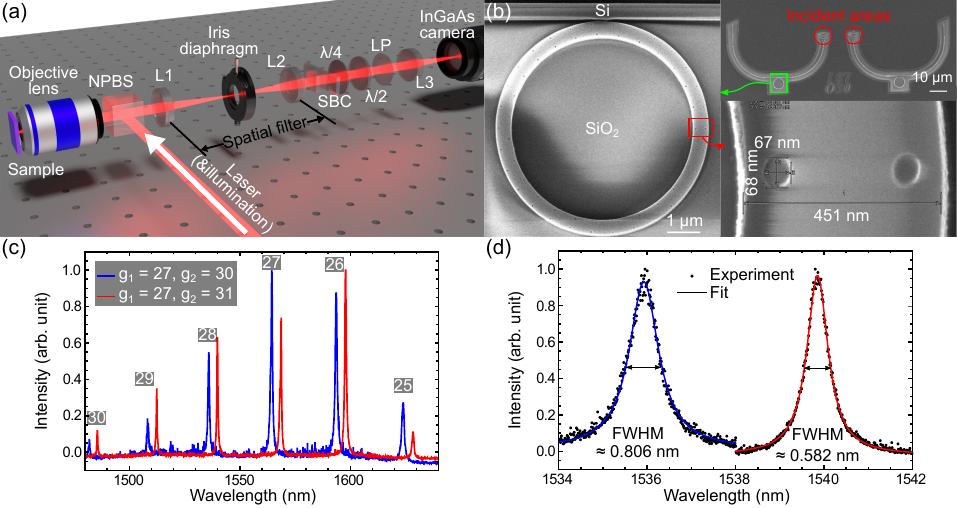}
    \caption{%
        (a)~An abstracted sketch of the optical measurement setup. L1, L2, and L3 are lenses. NPBS is a non-polarizing beam splitter, SBC is a Soleil-Babinet compensator, $\lambda/4$ and $\lambda/2$ are quarter and half waveplates, and LP is a linear polarizer. 
        (b)~Scanning electron microscope images of a fabricated device showing spectra indicated in blue lines in the following panels. A bus waveguide with one side terminated by a grating coupler is attached to a ring resonator, and its cavity mode is excited by a laser incident normal to the device plane.
        (c)~Spectra of two representative devices designed to generate first- and second-order skyrmionic beams indicated by blue and red solid lines, respectively. The spectral range corresponds to the operating range of the wavelength tunable laser used for excitation. The number superimposed on each resonance peak indicates the corresponding azimuthal order.
        (d)~Fine spectra within a range of about \SI{4}{\nano\meter} for the 28th-order modes. Lorentz function fits are overlaid on the raw data and are indicated by solid lines in the colors corresponding to (c).
    }
    \label{fig:OpticalCharacterization}
\end{figure*}
Figure~\ref{fig:OpticalCharacterization}(a) illustrates the optical setup used to inject laser light into the fabricated resonator and perform far-field polarimetry on the diffracted light. The incident laser is reflected by a non-polarizing beam splitter and directed onto the grating coupler couples with each ring resonator (shown in Fig.~\ref{fig:OpticalCharacterization}(b)) by a 50X objective lens (focal length $f=3.6$\,\si{\milli\meter}, numerical aperture $\mathrm{NA} = 0.65$). The beam diffracted from the device is collected by the same objective. After passing through the same beam splitter, the diffraction is trimmed from the incident spot by a spatial filter consisting of two lenses L1 ($f=200$\,\si{\milli\meter}) and L2 ($f=150$\,\si{\milli\meter}) and an iris diaphragm. A Soleil-Babinet compensator is inserted after the spatial filter to compensate for the unknown retardance of the optics preceding it. The light then sequentially passes through a quarter waveplate, half waveplate, and a linear polarizer to extract specific polarization components exclusively. The transmission axis of the polarizer is fixed to the normal direction of the optical table to negate polarization dependence of subsequent optics. Another lens L3 ($f=150$\,\si{\milli\meter}) is further inserted when converting the optical signal to the far field. Lenses L1, L2, and L3 are all arranged to focus on the back focal plane of their preceding lens. The optical signals are finally detected by a two-dimensional InGaAs camera (pixel size $20 \times 20$\,\si{\micro\meter^2}, equipped with an imaging lens of $f = 200$\,\si{\milli\meter}). Four images are captured by rotating the fast axes of the waveplates to correspond to four different polarization filtering setups: $S_0-S_1$, $S_0+S_1$, $S_0+S_2$, and $S_0-S_3$. The Stokes parameters $S_0$, $S_1$, $S_2$, and $S_3$ are then reconstructed from these four images.

Figure~\ref{fig:OpticalCharacterization}(c) shows the spectra of two resonators with different double grating configurations under clockwise excitation. The blue solid line represents the spectrum of the device depicted in Fig.~\ref{fig:OpticalCharacterization}(b), which features an angular grating with $g_1 = 27$ elements near the inner wall of the ring and $g_2 = 30$ elements near the outer wall. The red solid line represents another device with a similar design but with $g_2$ increased from 30 to 31, resulting in a beam with a different skyrmion number. Spectra are measured by integrating the intensity of the diffracted light while sweeping the incident laser wavelength. The free spectral range in the displayed wavelength region is approximately \SI{28}{\nano\meter}, and the $Q$ factors of the resonance modes range from 1000 to 5000, consistent with FDTD simulations. In these simulations, we assume nanoholes have a diameter of \SI{60}{\nano\meter} and a depth of \SI{110}{\nano\meter} (half the waveguide height, approximating the fabricated devices; see SM for details). Since simulations show no significant difference in resonance wavelength between $g_2=30$ and $31$, the observed red-shifted spectrum for $g_2=31$ is likely due to fabrication imperfections such as sidewall roughness and etching depth variations. Figure~\ref{fig:OpticalCharacterization}(d) displays detailed spectra for the $m=28$ WGMs, the target mode that operates within the telecommunication C-band. The Lorentzian function fits (indicated by the solid lines) yield $Q \approx 1900$ and $\approx 2600$ respectively for the $g_2=30$ and $31$ devices. The corresponding simulated $Q$ value is $\approx 4500$ for our original design (hole diameter of the grating elements is \SI{60}{\nano\meter}), and drops to $\approx 3000$ when the hole diameter of the grating elements is enlarged to \SI{70}{\nano\meter}. This value is closer to the measured value of the fabricated device, as shown in the bottom-right panel of Fig.~\ref{fig:OpticalCharacterization}(b). The absence of clear mode splitting in these spectra (in contrast to cases where the azimuthal order $m$ matches the grating periods $g_1$ or $g_2$, see SM) suggests that the backscatter rate due to angular gratings or other factors is relatively low compared to the photon decay rate of the $m=28$ mode.

\begin{figure}[htbp]
    \centering
    \includegraphics{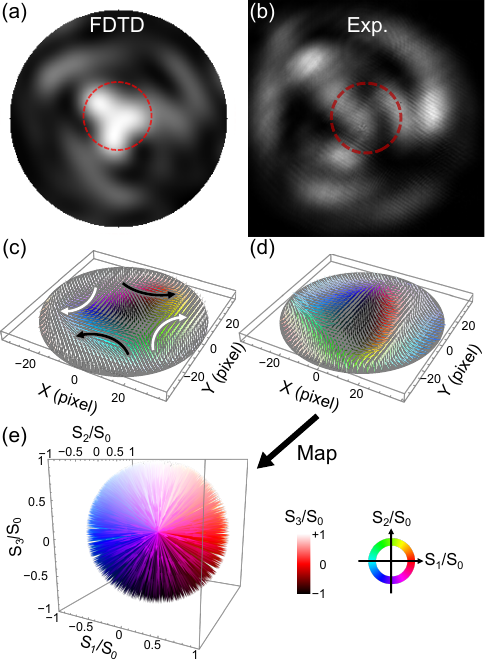}
    \caption{
        Far-field radiation of the resonance mode that is indicated by the blue line in Fig.~\ref{fig:OpticalCharacterization}(d). An anti-skyrmion with an $N_\mathrm{sk} = 1$ is expected to emerge.
        (a)~Simulated and (b)~experimentally observed far-field intensity profiles. In (a), a region of $\mathrm{NA}=0.65$, the value of the objective lens used in our optical setup, is showcased.
        (b)~Polarization distribution rendered by normalized Stokes vectors within the area enclosed by a red-dashed line in (a), which is the theoretically predicted domain in which an anti-skyrmion of $N_\mathrm{sk}=1$ emerges, and (d)~that of (b).
        (e)~A mapping of vectors in (d) onto the unit Poincar\'{e} sphere. Vectors pointing to all points on the sphere are included.
    }
    \label{fig:FirstOrder}
\end{figure}
Figures~\ref{fig:FirstOrder}(a) and (b) respectively show the simulated and the experimentally observed far-field radiation profiles of the device with a grating arrangement of $g_1 = 27$ and $g_2 = 30$, corresponding to the spectrum indicated by the blue line in Fig.~\ref{fig:OpticalCharacterization}(d). In this design, the inner angular grating diffracts $\ket{\circlearrowright}$-polarized light with an OAM order of $l_1=-(28-27)+1=0$ for the 28th-order clockwise WGM, while the outer one diffracts $\ket{\circlearrowleft}$-polarized light with an OAM order of $l_2=-(28-30)-1=1$. Figures~\ref{fig:FirstOrder}(c) and (d) present the evaluated polarization profiles of the total diffraction, visualized by the normalized Stokes vectors $\boldsymbol{s}$ in the region enclosed by the red dashed line in Figs.~\ref{fig:FirstOrder}(a) and (b), respectively. This region is the theoretically predicted domain in which a target skyrmion emerges, which is an area with a radius of $\approx37$ pixels (corresponding to a numerical aperture $\mathrm{NA} \approx 0.2$) in the InGaAs area detector used in our optical setup. Unlike Fig.~\ref{fig:Concept}(a), which is computed based on a theoretical toy model calculation, both the intensity and polarization show triangularly deformed distribution rather than an axisymmetric profile. This discrepancy is likely due to imperfect circular polarization generated by the angular gratings, a topic we will discuss later. Figure~\ref{fig:FirstOrder}(e) maps the Stokes vectors from (d) to the unit Poincar\'{e} sphere. Since the sphere is fully covered, all polarization states are expected to be represented within this region, fulfilling the skyrmionic topology.

For qualitatively characterizing the skyrmion topology, we can decompose skyrmion number $N_\mathrm{sk}$ into the product of two quantities: the polarity $P$ and vorticity $M$~\cite{nagaosa2013topological}. The polarity is half of the difference in the out-of-plane component ($S_3/S_0$) between the edge and the center of the vector field. The vorticity is the winding number of the in-plane vector component, equal to the difference in OAM. The simulated and experimentally observed radiations are $\ket{\circlearrowright}$-polarized ($S_3/S_0 = -1$) at the beam center and $\ket{\circlearrowleft}$-polarized ($S_3/S_0 = 1$) near the domain boundary, resulting a polarity of $P = (-1 - (+1))/2 = -1$. The vorticity is given by the difference in OAM $M = l_1-l_2=-1$, which appears as the antivortex (saddle-like) distribution of the in-plane component of the Stokes vectors. Accordingly, this light beam is characterized as $N_\mathrm{sk} = P M = +1$. In this paper, we adopt the definition of anti-skyrmion as having a negative vorticity $M$, rather than a negative skyrmion number\cite{koshibae2016theory,guo2020meron}. The polarization texture is therefore an optical anti-skyrmion of $N_\mathrm{sk} = +1$. In addition to qualitative discussions, this topological number can be numerically verified by computing the skyrmion number density distribution based on Eq.~(\ref{eq:1}) and integrating it over the target skyrmion domain. Our experimental results yield $N_\mathrm{sk} = 1.01$ for a 37-pixel radius domain and $1.00$ for a 38-pixel domain, in good agreement with theoretical predictions.

\begin{figure}[htbp]
    \centering
    \includegraphics{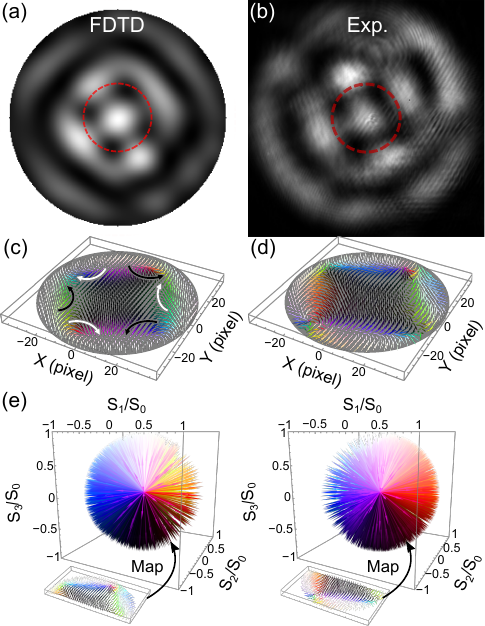}
    \caption{
        Far-field radiation of the resonance mode that is indicated by the red line in Fig.~\ref{fig:OpticalCharacterization}(d). n anti-skyrmion with an $N_\mathrm{sk} = 2$ is expected to emerge.
        (a)~Simulated and (b)~experimentally observed far-field intensity profiles. In (a), a region of $\mathrm{NA}=0.65$ is showcased.
        (c)~Polarization distribution rendered by normalized Stokes vectors within the area enclosed by a red-dashed line in (a), which is the theoretically predicted domain in which an anti-skyrmion of $N_\mathrm{sk}=2$ emerges, and (d)~that of (b).
        (e)~Mappings of vectors in half areas of (d) onto respective unit Poincar\'{e} spheres. Each half possesses vectors pointing to all points on the sphere.
    }
    \label{fig:SecondOrder}
\end{figure}
Figure~\ref{fig:SecondOrder} shows the far-field radiation profile and polarization texture of the mode indicated by the red line in Fig.~\ref{fig:OpticalCharacterization}(d). The increased grating period, $g_2 = 31$, leads the corresponding angular grating to diffract a light beam with an increased OAM order of $2$ for the $\ket{\circlearrowright}$-polarized component, while the OAM order of the $\ket{\circlearrowleft}$-polarized component remains zero due to the unchanged $g_1$ configuration. Thus, the vorticity becomes $-2$ while the polarity doesn't change from $-1$, resulting in a skyrmion number of $+2$.
The areas surrounded by the red dashed-line in Figs.~\ref{fig:SecondOrder}(a) and (b) indicate a 37-pixel radius domain, similar to those in Figs.~\ref{fig:FirstOrder}(a) and (b). The polarization texture within this domain is rendered in Fig.~\ref{fig:SecondOrder}(c) and (d), respectively. Since a round trip along the azimuthal direction encounters vectors with identical orientations twice, the field has second-order vorticity. Additionally, dividing the domain into two parts and mapping the vector field onto the unit Poincaré sphere shows that each part fully wraps the sphere, indicating the unique nature of a second-order anti-skyrmion. The computed skyrmion number for the presented vector field, based on Eq.~(\ref{eq:1}), is $2.00$, supporting the generation of a second-order optical anti-skyrmionic beam.

While the skyrmion numbers obtained from far-field polarimetry are close to theoretical predictions, the intensity and polarization distributions differ from those predicted by the theoretical toy model described in \cite{lin2021microcavity}. The model predicts axisymmetric intensity distribution and concentric polarization distribution, as shown in Fig.~\ref{fig:Concept}(a). However, the distributions observed in Fig.~\ref{fig:FirstOrder} and Fig.~\ref{fig:SecondOrder} appear to be ``triangulated'' and ``squared'', respectively. These discrepancies may arise from backscattering due to diffraction gratings and imperfections in circular polarization generated by these gratings, which were not considered in the previous toy model. Extending the toy model to include these effects reveals that the observed errors are primarily due to imperfect circular polarization (see SM for model details). Therefore, a more ideal optical skyrmion could be achieved by developing techniques for fabricating shallow holes with smaller radii to improve the degree of circular polarization. Notably, while imperfections in circular polarization can perturb the skyrmionic topology, backscattering alone does not destroy the skyrmion topology.

\section{Discussions}
In the preceding results, we have presented the outcomes of driving a clockwise WGM. Upon driving the corresponding counterclockwise WGM, both the OAM and circular polarization undergo inversion of handedness. Consequently, the polarization texture obtained shows a reversal in the sign of the polarity, while the vorticity remains unchanged. According to our definition, this results in maintaining the type of skyrmion (skyrmion or anti-skyrmion), but inverting the sign of the skyrmion number. We fabricated devices capable of exciting a counterclockwise WGM and confirmed that the polarity of the far-field polarization distribution is indeed inverted (see SM). However, due to potential fabrication imperfections, these devices exhibit slightly deviated skyrmion numbers compared to their counterparts operating in the clockwise mode ($-0.84$ and $-1.73$ for design values of $-1$ and $-2$, respectively).

As we have demonstrated experimentally, the skyrmion number can easily be controlled in straightforward manners, either by altering the number of grating periods or by switching the handedness of the driven WGM, akin to methods used for on-chip OAM lasers~\cite{cai2012integrated,miao2016orbital,zhang2018inp,shao2018spin,massai2022pure}. Besides perturbations along the chiral lines in the ring waveguide discussed here, perturbing the sidewalls of the ring waveguide, as employed in most OAM emitters, may satisfy the same objective. Nevertheless, arranging diffraction gratings on the ring, rather than on their sidewalls, may be less susceptible to perturbations from neighboring optical structures such as bus waveguides, thus offering higher design flexibility.

In addition to the passive skyrmionic beam generator based on the silicon photonic platform we demonstrated, it is also feasible to implement a stand-alone microcavity capable of generating an optical skyrmion laser by incorporating a gain medium, potentially using III-V compound semiconductors, similar to those in OAM micro lasers~\cite{miao2016orbital,zhang2018inp,hayenga2019direct}. Furthermore, skyrmionic single photons could be generated by embedding a single-photon source, such as semiconductor quantum dots, in the ring waveguide. The ring resonator structure with double diffraction gratings proposed here thus presents a promising platform for realizing optical quantum states with intricate spin and orbital angular momentum couplings.

\section{Conclusions}
In this study, we have demonstrated microcavity-based generations of light beams in which polarization textures are characterized by the second homotopy group of spheres, i.e., skyrmions. By employing dual angular gratings composed of shallow airholes arranged on a microring cavity, we achieved independent control of the orbital angular momentum for both left and right-handed circular polarizations. This approach enables the synthesis of beams with polarization textures defined by first and second-order anti-skyrmions, which can be achieved by precisely tuning the OAM for each circularly polarized beam individually. Moreover, our results highlight the versatility of our approach in generating anti-skyrmions with tailored topological numbers by simply adjusting the number of diffraction grating periods.

Building upon the methodology demonstrated in this work, integrated optical skyrmion lasers can be realized by incorporating gain media with a non-reciprocal emission mechanism, akin to existing OAM lasers~\cite{miao2016orbital,hayenga2019direct,lin2019spin,dai2024rate}. Such devices hold promise for applications in optical telecommunications similar to OAM multiplexing telecommunications. Moreover, when combined with quantum emitters such as quantum dots, they could find utility in quantum information processing. Extending our approach to superimpose additional spatial modes could enable the realization of higher-order topological structures, such as optical hopfions~\cite{sugic2021particle,ehrmanntraut2023optical,shen2023topological}. Spatiotemporal topologies, recently demonstrated in polychromatic light~\cite{chong2020generation,alonso2020complete,wan2022toroidal}, may also be generated by integrating multiple angular-graded microring resonators with different resonance wavelengths. The unique polarization heterogeneity created by our approach also holds potential applications in fields beyond photonics, such as atomic physics. In this context, light beams with exotic spatial structures now have found potential applications~\cite{hattori2022integrated,parmee2022optical}, and thus, methods to generate complex structured light using nanophotonic technologies are highly demanded. Our on-chip skyrmionic beam generators are promising candidates for integrating structured light into modularized atomic systems.

\acknowledgments{This research was supported by JST, Core Research for Evolutionary Science and Technology (CREST) Grant Number JPMJCR19T1, Japan.}

\bibliography{ms}

\end{document}


\title{Supplementary Material for On-Chip Optical Skyrmionic Beam Generators}
\maketitle

\renewcommand{\thesection}{S\arabic{section}}
\renewcommand{\theequation}{S\arabic{equation}}
\renewcommand{\thefigure}{S\arabic{figure}}
\setcounter{secnumdepth}{1}
\renewcommand{\bibnumfmt}[1]{[S#1]}
\renewcommand{\citenumfont}[1]{S#1}

\section{Device fabrication}
The nanostructures are first patterned on an electron beam resist (ZEP520A) that is spin-coated onto a silicon-on-insulator substrate via electron beam lithography.
The nanopatterns in the electron beam resist are then transferred to the silicon layer through reactive ion etching with an SF$_6$/O$_2$ mixture. Optical waveguides, including microrings and diffraction gratings on these rings, are fabricated in a single lithography step to minimize alignment errors. The etching process is carefully controlled to minimize over-etching for larger features, such as waveguides. The grating elements, having much finer dimensions, exhibit a slower etching rate, leading to a shallower etch. For these fine patterns, the electron beam dosage is adjusted compared to other patterns, resulting in an etching depth approximately half that of the silicon-on-insulator layer. This specific etch depth ensures that the diffracted light exhibits a high degree of circular polarization.

\section{Depth of grating elements}
\begin{figure}[htbp]
    \centering
    \includegraphics{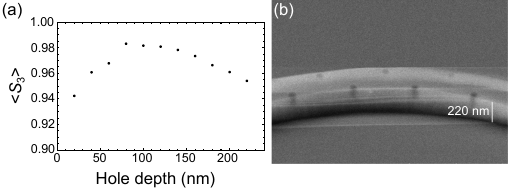}
    \caption{%
        (a)~Simulated dependence of the expected value of the degree of circular polarization $\braket{S_3}$ on the depth of angular gratings. The silicon ring waveguide has a height of \SI{220}{\nano\meter} and the radius of the grating elements is \SI{30}{\nano\meter}. The degree of circular polarization reaches its maximum at approximately half the height of the waveguide.
        (b)~A bird's-eye view SEM micrograph of a fabricated device.  The grating elements (nano air holes) are positioned close to the sidewall of the ring waveguide and appear as shadows in the secondary electron image. The depth of the nanoholes (with a design radius of \SI{30}{\nano\meter}) is about half the height of the waveguide.
    }
    \label{fig:HoleDepth}
\end{figure}

In addition to the positioning of the grating elements (nano air holes), their depth also impacts the polarization state of the diffracted light. Figure~\ref{fig:HoleDepth}(a) showcases the influence of hole depth on the degree of circular polarization, as revealed by FDTD simulations. These simulations are based on a ring cavity structure with 31 grating elements positioned near the outer edge, which ideally diffract light in a state of $\ket{\circlearrowleft,+2}$ when the 28th-order clockwise WGM is excited. The polarization of the far-field radiation is characterized by computing the expected value of a normalized Stokes parameter:
\begin{equation} \braket{S_3} = \frac{\braket{E|\sigma_{2}|E}}{\braket{E|E}}, \end{equation}
where $\ket{E}$ represents the electric field vector of the diffracted beam and $\sigma_{2}$ is a Pauli matrix. The integration is performed over a region within a divergence angle $\theta \leq 12^\circ$ (numerical aperture $\mathrm{NA} \approx 0.2$), which is the domain where the target skyrmion is expected to be observed. The degree of circular polarization is maximized when the depth of the hole is slightly less than half the height of the waveguide. The devices we have fabricated approximately meet this condition, as depicted in Fig.~\ref{fig:HoleDepth}(b).

\section{Counter-clockwise WGM excitation}
\begin{figure}[htbp]
    \centering
    \includegraphics{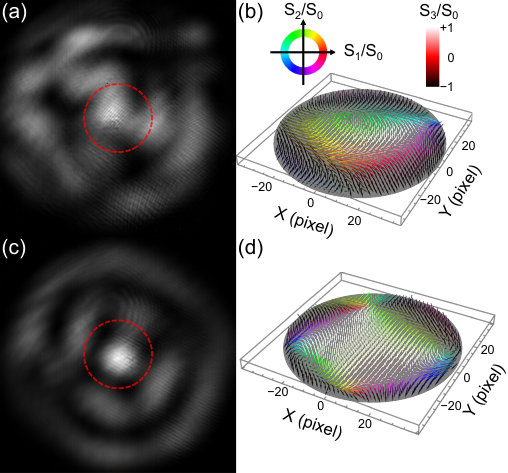}
    \caption{%
        Far-field (a)~intensity profile and (b)~Stokes vector distribution of a device with a double angular grating configuration of $g_1 = 27$ and $g_2 = 30$. A counterclockwise WGM is excited by a laser with a wavelength of \SI{1540.08}{\nano\meter}. The red dashed line in (a) indicates a 37-pixel radius circle, , which corresponds to the visualized domain shown in (b).
        (c--d)~Same set of plots as (a--b) but for a device with a double angular grating configuration of $g_1 = 27$ and $g_2 = 31$, under counterclockwise excitation at a wavelength of \SI{1536.63}{\nano\meter}.
    }
    \label{fig:CCW}
\end{figure}

Figures~\ref{fig:CCW}(a--b) show the far-field intensity profile and polarization structure for a device with the same cavity design ($g_1 = 27$ and $g_2 = 30$) as in Fig.~3, but with a different bus waveguide design to excite counterclockwise WGM instead of clockwise WGM. The resulting polarization profile differs from that in Fig.~3(b), featuring $\ket{\circlearrowleft}$-polarized light at the center of the beam and $\ket{\circlearrowright}$-polarized light at the domain boundary, thus inverting the polarization. The expected skyrmion number for this configuration is $-1$, while the experimentally obtained value is $-0.84$, likely due to fabrication imperfections. Another indication that this ring cavity is not structurally identical to the one in Fig.~3 is the center wavelength of resonance, which is \SI{1540.08}{\nano\meter} for this device, compared to \SI{1535.9}{\nano\meter} for the device in Fig.~3. Figures~\ref{fig:CCW}(c--d) present similar data for a counterpart to the device in Fig.~4. For this device, the observed skyrmion number is $-1.73$, while the design value is $-2$, and the resonance wavelength is \SI{1536.63}{\nano\meter}, in contrast to \SI{1539.78}{\nano\meter} for the device in Fig.~4.

\section{Large backscattering upon synchronous mode order and lattice periods}
\begin{figure}[htbp]
    \centering
    \includegraphics{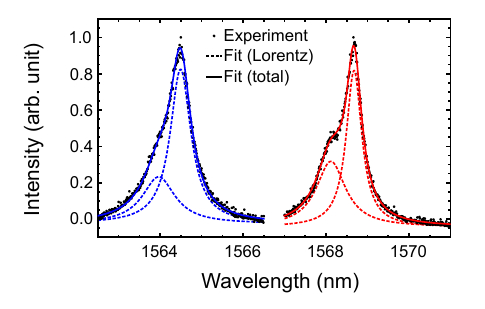}
    \caption{%
        Spectra of the cavity modes for the 27th azimuthal order. The spectra are not single-peaked and are fitted by two Lorentzian functions (indicated by the dashed lines). The solid lines represent the sum of these two Lorentzian fits.
    }
    \label{fig:m27}
\end{figure}

When the azimuthal order of the WGM $m$ matches one of the two grating periods, $g_1$ or $g_2$, reflections at each period are in phase, leading to a large backscatter. Figure~\ref{fig:m27} presents detailed spectra for the $m=27$ mode shown in Fig.~2(b), where $m=g_1$. The spectra are distinctly not unimodal and are fitted by two Lorentzian functions. Since the spectra are not fully split in these devices, the backscatter rate is comparable to the photon decay rate.

\section{Discussion on the effect of backscattering on the skyrmionic polarization texture}
In our previous report~\cite{lin2021microcavity}, our toy model assumes perfectly unidirectional WGM, either clockwise or counter-clockwise. However, in actual devices, finite reflections and the excitation of counter-rotating WGMs due to fabrication imperfections, diffraction gratings, and auxiliary structures such as the bus waveguide introduce complexities. In this section, we extend our model to analyze how counter-rotating WGMs impact skyrmionic topologies.

When a double grating diffracts a superposition of two angular momentum states, $\ket{\circlearrowleft,l_1}$ and $\ket{\circlearrowright,l_2}$ the angular momentum of the diffracted light changes to $\ket{\circlearrowright,-l_1}$ and $\ket{\circlearrowleft,-l_2}$ for the counter-rotating WGM with the same azimuthal order. The superposition of these four angular momentum states, including the counter-rotating WGM, alters the polarization distribution compared to the case where the counter-rotating component is absent. Nevertheless, one can still find a skyrmionic topology in the cross-sectional polarization distribution of the superimposed beam.

Consider a simple case with $l_1 = 0$ and $l_2 = l \neq 0$. The complex amplitude of the diffracted light from the counter-clockwise WGM can be expressed as
\begin{equation}
    \psi_\mathrm{CCW}\left( r,\phi \right) =
    A_1 J_{0}\left( k_1 r \right)\ket{\circlearrowleft} + 
    i^{-\left|l\right|}A_2 J_{\left| l \right|}\left( k_2 r \right) e^{i l \phi} \ket{\circlearrowright},
    \label{eq:farFieldElectricFieldProfile}
\end{equation}
where $k_1 = \frac{2\pi}{\lambda}\rho_1$ and $k_2 = \frac{2\pi}{\lambda}\rho_2$.
Here, $\rho_1$ and $\rho_2$ are the radii of the inner and outer circular gratings, respectively, and $A_1$ and $A_2$ are the amplitudes of the diffraction contributions from the inner and outer gratings. For simplicity, we assume $A_1$ and $A_2$ are real, indicating that the two diffractions are in phase at $\phi=0$. The contribution from the clockwise WGM is given by 
\begin{equation}
    \psi_\mathrm{CW}\left( r,\phi \right) =
    R\left(
    A_1 J_{0}\left( k_1 r \right)\ket{\circlearrowright} + 
    i^{-\left|l\right|}A_2 J_{\left| l \right|}\left( k_2 r \right) e^{- i l \phi} \ket{\circlearrowleft}
    \right),
    \label{eq:farFieldElectricFieldProfileWithBackscattering}
\end{equation}
where $R$ represents the factor of CW mode excitation relative to the CCW mode. Again, for simplicity, we assume $R$ is a real number with $\left|R\right| \leq 1$. The skyrmion number density (or the Berry curvature) is then given by
\begin{equation}
    dN_\mathrm{sk}\left(r, \phi\right) = l\frac{
        A_1 {A_2}^2 J_{\left|l\right|}\left(k_2 r\right) \alpha\left(r\right) \beta\left(r, \phi\right)
    }{2\pi \left(
        \left(1+R^2\right) \left({A_1}^2 J_0\left(k_1 r\right)^2+{A_2}^2 J_{\left|l\right|}\left(k_2 r\right)^2\right)
        + 4 R A_1 A_2 \cos\left(l\frac{\pi}{2}\right) \cos \left(l \phi \right) J_0(k_1 r) J_{\left|l\right|}\left(k_2 r\right)
    \right)^2},
    \label{eq:berryCurvatureWithBackscattering}
\end{equation}
where
\begin{equation}
    \alpha\left(r\right) = 2 k_1 J_1\left(k_1 r\right) J_{\left|l\right|}\left(k_2 r\right) + k_2 J_0\left(k_1 r\right) \left( J_{\left|l-1\right|}\left(k_2 r\right) - J_{\left|l+1\right|}\left(k_2 r\right)\right) \nonumber
\end{equation}
and
\begin{equation}
    \beta\left(r, \phi\right) = \left(1-R^4\right) A_1 J_0(k_1 r) - 2 R A_2 J_{\left|l\right|}(k_2 r) \left(R^2 \cos \left(l\left(\frac{\pi}{2} - \phi\right)\right)-\cos \left(l\left(\frac{\pi}{2} + \phi\right)\right)\right) \nonumber.
\end{equation}
Equation~(\ref{eq:berryCurvatureWithBackscattering}) reveals three boundary conditions where the sign of the Berry curvature may flip: $J_{\left|l\right|}\left(k_2 r\right) = 0$, $\alpha\left(r\right) = 0$, or $\beta\left(r, \phi\right) = 0$. The first condition corresponds to a circular loop where the $\left|l\right|$-th Bessel function vanishes, and the polarization state matches that at the center of the beam cross-section. The second condition corresponds to the scenario where the radial derivative of the ratio of diffraction by the two gratings, $\propto \frac{\partial}{\partial r}\frac{J_{\left|l\right|}\left(k_2 r\right)}{J_0(k_1 r)}$, is zero, indicating that the Berry curvature is zero because the polarization state is preserved. The third condition gives an iso-polarization loop of normalized Stokes vectors distinct from the beam center:
\begin{equation}
    \boldsymbol{s} = \left( (-1)^{l-1} \frac{2R}{1+R^2}, 0, -\frac{1-R^2}{1+R^2}\right),
\end{equation}
which corresponds to the zero-OAM component $J_{0}\left(k_1 r\right) = 0$ contributing exclusively to the polarization state at the center when $R = 0$. When the OAM order $l$ is even, this iso-polarization loop is in a polarization state exactly orthogonal to that at the beam center. When $l$ is odd, the iso-polarization loop is orthogonal to the polarization state at a point where $r$ satisfying $J_{\left|l\right|}(k_2 r) = \mathrm{sgn}\left(l\right)\frac{2\left|R\right|}{1-R^2}\frac{A_1}{A_2}J_0(k_1 r)$ and $\phi=-\mathrm{sgn}\left(R\right)\frac{\pi}{2}$. Here, $\mathrm{sgn}$ is the sign function. Let $r_0$ be the radius where $J_0(k_1 r)$ first becomes zero. This point will initially appear within $r < r_0$, while $\beta\left(r, \phi\right) = 0$ leads to $J_{\left|l\right|}(k_2 r) = -\mathrm{sgn}\left(l\right)\frac{1-R^2}{2\left|R\right|}\frac{A_1}{A_2}J_0(k_1 r)$, which implies $r \geq r0$. Consequently, there will always be a point where the polarization state is orthogonal to that at the boundary within the region enclosed by the first iso-polarized closed loop defined by $\beta\left(r, \phi\right) = 0$. The skyrmion number calculated in this region is $N_\mathrm{sk}=l$ when $R=0$, and this value remains consistent even when $R \neq 0$.

\begin{figure}[htbp]
    \centering
    \includegraphics{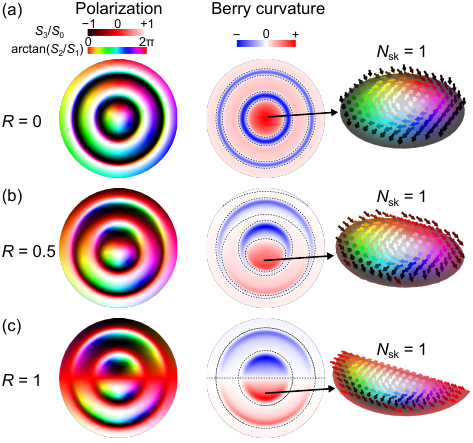}
    \caption{%
        (a)~Polarization distribution in the far field at $R = 0$ (left panel) and Berry curvature profile derived from the corresponding normalized Stokes vectors (middle panel) calculated based on our toy model (Eqs.~(\ref{eq:farFieldElectricFieldProfile}) and (\ref{eq:farFieldElectricFieldProfileWithBackscattering})) with parameters $l=1$, $\rho_1 = 3-0.15$\,\si{\micro\meter}, $\rho_2 = 3+0.15$\,\si{\micro\meter}, $\lambda=1.54$\,\si{\micro\meter}, and $A_1 = A_2 = 1$. The area corresponding to $r \leq 0.65$ (or $\mathrm{NA}=0.65$) is rendered. Dashed lines indicate the zeros of the Berry curvature, and the distribution of Stokes vectors within the most central region enclosed by the dashed lines is shown in the right panel.
        (b) and (c) show the same set of plots for $R=0.5$ and $R=1$, respectively.
    }
    \label{fig:Backscatter}
\end{figure}
Figures~\ref{fig:Backscatter}(a--c) showcase the distributions of polarization states and Berry curvature profiles calculated from the model for $l=1$. Parameters such as $\rho_1$, $\rho_2$, and $\lambda$ are taken from our device design. When $R=0$, concentric domains of positive and negative Berry curvature emerge alternately, as shown in Fig.~\ref{fig:Backscatter}(a). The Stokes vector field exhibits a nested structure of skyrmions with $N_\mathrm{sk} = +1$ and $-1$ alternating within the beam cross-section. At the boundary of the most central skyrmion, the normalized Stokes vector heads exactly downward. As exemplified in Fig.~\ref{fig:Backscatter}(b), when a finite counter-rotating WGM is present, the boundary is deformed from a perfect circle while maintaining the nested skyrmion structure, and the Stokes vectors at the boundary are identical but fall towards the in-plane direction. If the CW and CCW WGMs are balanced and form a standing wave mode, the Stokes vectors are oriented completely in-plane at the boundary, forming an in-plane skyrmion~\cite{moon2019existence}.

This discussion indicates that even with the excitation of a counter-rotating WGM due to backscattering or other factors, the diffracted light beam will retain a skyrmionic topology. The reflected power within the ring, calculated from the power coupled from the ring resonator to the bus waveguide in FDTD simulations, is typically less than 1\%.

\section{Discussion of the effect of imperfections in the circular polarization of diffraction on the skyrmionic topology}
n addition to backscattering, another potential discrepancy between our theoretical model and experimental observations is the degree of circular polarization in the diffracted light. Our theoretical model assumes perfect circular polarization of the diffracted light. However, in practical devices, the circular polarization can be imperfect, as shown in Fig.~\ref{fig:HoleDepth}(a). This imperfection may result from factors such as misalignment of the angular gratings with the chiral lines or the finite size of the grating elements. When the diffracted light is not perfectly circularly polarized, additional components $\ket{\circlearrowright,l_1+2}$ and $\ket{\circlearrowleft,l_2-2}$ are introduced to the angular momentum states $\ket{\circlearrowleft,l_1}$ and $\ket{\circlearrowright,l_2}$, respectively.
Assuming the ideal diffracted light is described by Eq.~(\ref{eq:farFieldElectricFieldProfile}), imperfections in circular polarization can be represented by adding
\begin{eqnarray}
    \psi_\mathrm{CCW'}\left( r,\phi \right) = &-\beta_1 A_1 J_{2}\left( k_1 r \right) e^{i 2 \phi} \ket{\circlearrowright} + i^{-\left|l-2\right|} \beta_2 A_2 J_{\left| l - 2 \right|}\left( k_2 r \right) e^{i \left(l-2\right) \phi} \ket{\circlearrowleft},
    \label{eq:farFieldElectricFieldProfileWithPolarizationImperfection}
\end{eqnarray}
where $\beta_1$ and $\beta_2$ are real numbers quantifying the deviation of the polarization state from perfect circular polarization. When $\beta_1$ and $\beta_2$ are non-zero, we generally do not observe an iso-polarization loop as we do in the case with perfect circular polarization.

\begin{figure}[htbp]
    \centering
    \includegraphics{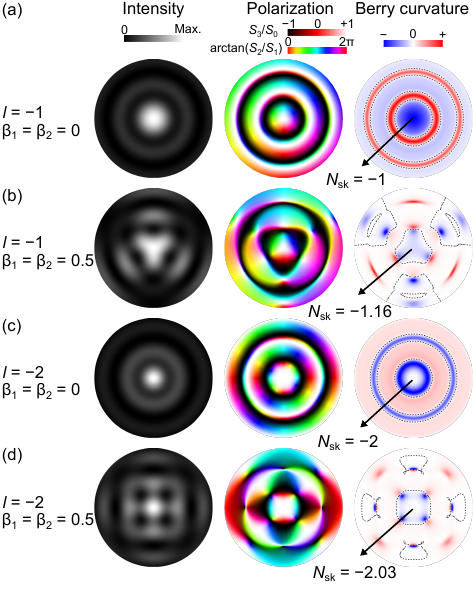}
    \caption{%
        (a)~Far field intensity (left panel), polarization (middle panel), and Berry curvature (right panel) distributions computed based on our toy model (Eqs.~(\ref{eq:farFieldElectricFieldProfile}) and (\ref{eq:farFieldElectricFieldProfileWithPolarizationImperfection})). Parameters are $l=-1$, $d_1 = 3-0.15$\,\si{\micro\meter}, $d_2 = 3+0.15$\,\si{\micro\meter}, $\lambda=1.54$\,\si{\micro\meter}, $A_1 = A_2 = 1$, and $\beta_1 = \beta_2 = 0$. The rendered area corresponds to $r \leq 0.65$ (or $\mathrm{NA}=0.65$). Dashed lines in the right panel indicate contours where Berry curvature is zero.
        (b--d)~Same sets of plots to (a) but for different conditions: (b)~$l=-1$, $\beta_1 = \beta_2 = 0.5$; (c)~$l=-2$, $\beta_1 = \beta_2 = 0$; (d)~$l=-2$, $\beta_1 = \beta_2 = 0.5$.
    }
    \label{fig:ImperfectCircularPol}
\end{figure}
Figures~\ref{fig:ImperfectCircularPol}(a--d) exemplify the effect of imperfect circular polarization. Figure~\ref{fig:ImperfectCircularPol}(a) shows the analytical results when the light is ideally circularly polarized, based on the design corresponding to Figs.~\ref{fig:CCW}(a--b). The intensity and Berry curvature distributions are axisymmetric, and an anti-skyrmion with $N_\mathrm{sk} = -1$ appears at the beam center. Supposing that the degree of circular polarization is imperfect, results assuming $\beta_1 = \beta_2 = 0.5$ are shown in Fig~\ref{fig:ImperfectCircularPol}(b). The degree of circular polarization derived from Fig.~\ref{fig:HoleDepth}(a) suggests $\beta_2 \approx 0.15$. However, a value of about 0.5 better fits the experimental observations, indicating an additional factor that reduces the degree of circular polarization. Under these conditions, the distribution deviates from axisymmetry, reflecting the experimentally observed pattern in Figs.~\ref{fig:CCW}(a--b). This deviation is due to the addition of components with different OAM values, resulting in a three-fold rotational symmetry. As a result, while the beam center remains $\ket{\circlearrowleft}$-polarized, no $\ket{\circlearrowright}$-isopolarized loop is present around it. Although a region of negative Berry curvature can be identified around the beam center, the skyrmion number, derived from the integral of the Berry curvature over this domain, will be a non-integer and exceed the design value. 

Figures~\ref{fig:ImperfectCircularPol}(c--d) show similar results for a second-order anti-skyrmion. The four-fold rotational symmetry observed in Figs.~\ref{fig:CCW}(c--d) is reproduced when diffraction is imperfectly circularly polarized. These findings underscore that the degree of circular polarization in diffraction significantly influences the skyrmionic topology of the emitted light. Thus, improving the degree of circular polarization is crucial. Techniques that focus on fabricating nanoholes with smaller radii could enhance the circular polarization, thereby improving the skyrmionic topology of the generated beam.

\bibliography{sm}